\documentstyle[amssymb,prl,aps,multicol,psfighund]{revtex}
\renewcommand{\narrowtext}{\begin{multicols}{2}
\global\columnwidth20.5pc} 
\renewcommand{\widetext}{\end{multicols}
\global\columnwidth42.5pc} \multicolsep = 8pt plus 4pt minus 3pt

\begin{document}
\draft
\title{Direct Coulomb and Exchange Interaction in Artificial Atoms}
\author{S. Tarucha$^{1,2}$, D.G. Austing$^{2}$, Y. Tokura$^{2}$, W.G. van der Wiel$%
^{3}$, and L.P. Kouwenhoven$^{3}$}
\address{$^{1}$Department of Physics, University of Tokyo, 7-3-1, Hongo, Bunkyo-ku,
Tokyo 113-0033, Japan\\
$^{2}$NTT Basic Research Laboratories, 3-1, Morinosato Wakamiya, Atsugi-shi,
Kanagawa 243-0198, Japan\\
$^{3}$Department of Applied Physics and DIMES, Delft University of\\
Technology, PO Box 5046, 2600 GA Delft, The Netherlands}
\date{\today}
\maketitle

\begin{abstract}
We determine the contributions from the direct Coulomb and exchange
interactions to the total interaction in semiconductor artificial atoms. We
tune the relative strengths of the two interactions and measure them as a
function of the number of confined electrons. We find that electrons tend to
have parallel spins when they occupy nearly degenerate single-particle
states. We use a magnetic field to adjust the single-particle state
degeneracy, and find that the spin-configurations in an arbitrary magnetic
field are well explained in terms of two-electron singlet and triplet states.
\end{abstract}


\narrowtext

The addition of a single electron charge to a quantum box costs a certain
energy, which is responsible for Coulomb blockade in electron transport [1].
Also a change in spin is associated with a certain change in energy,
e.g.exchange energy is gained when electrons are added with parallel spins
as compared to anti-parallel spins. Depending on the system, a large total
spin (ferromagnetic filling) or a minimum total spin value
(anti-ferromagnetic filling) is favored. In semiconductor quantum dots
alternate spin filling [2] as well as spin-polarized filling [3] have been
reported. Here, we study vertical quantum dots which have well-defined
single-particle states. When these states are separated by a large energy, $%
\Delta E$, an anti-ferromagnetic filling is favored. For small $\Delta E$, a
ferromagnetic filling is observed, which is in line with Hund's first rule
from atomic physics. We use a magnetic field, $B$, to tune $\Delta E$($B$)
allowing us to alter the spin filling. 
We first discuss a simple model that describes filling of two
single-particle states with two interacting electrons. Fig. 1(b) shows two,
spin-degenerate single-particle states with energies $E_{a}$ and $E_{b}$
crossing each other at $B=B_{0}$. The ground state (GS) energy, $U$(1), for
one electron occupying these states, equals $E_{a}$ for $B$ $<$ $B_{0}$ and $%
E_{b}$ for $B>B_{0}$ (thick line in Fig. 1(b)). For two electrons we can
distinguish four possible configurations with either total spin $S$ $=0$
(spin-singlet) or $S$ = 1 (spin-triplet). (We neglect the Zeeman energy
difference between $S_{z}=-1,0$ and 1.) The corresponding energies, $U_{i}$%
(2, $S$) for $i=1$ to 4, are given by: $%
U_{1}(2,0)=2E_{a}+C_{aa},U_{2}(2,0)=2E_{b}+C_{bb}$, $%
U_{3}(2,1)=E_{a}+E_{b}+C_{ab}-|K_{ab}|$, $U_{4}(2,0)=E_{a}$ $+E_{b}+C_{ab}$.
Here, $C_{ij}$ $(i,j=a,b)$ is the direct Coulomb (DC) energy between two
electrons occupying states with energies $E_{i}$ and $E_{j}$, and $K_{ab}$
is the exchange (EX) energy ($K_{ab}$ $<0$) between two electrons occupying $%
E_{a}$ and $E_{b}$ with parallel spins [4]. 
The experiments below measure the electrochemical potential defined for a
two electron system as $\mu (2)\equiv U(2)-U(1)$. For each $U_{i}(2)$ we
obtain the potentials: $\mu _{i}(2)=U_{i}(2)-E_{a}$ for $B$ $<$ $B_{0}$ and $%
\mu _{i}(2)=U_{i}(2)-E_{b}$for $B$ $>B_{0}$ (see Fig. 1(c)).  
\begin{figure} 
\centerline{\psfig{figure=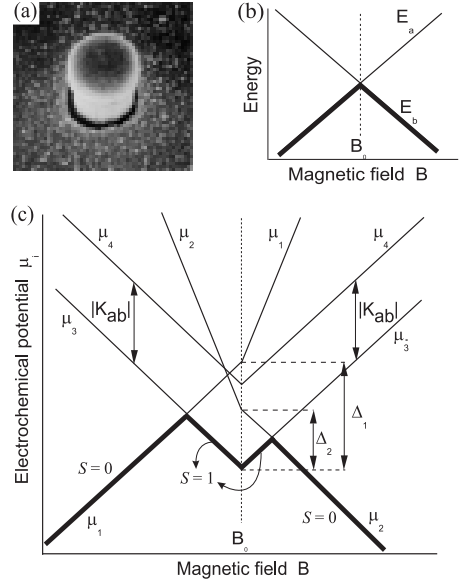,width=8.5cm,height=11cm,rheight=8cm}}
\caption{(a) Scanning electron micrograph of the semiconductor quantum dot device. The
quantum dot is located inside the  0.5 $\mu $m-diameter pillar.
The top and bottom contacts serve as source and drain electrodes. A Schottky gate is wrapped
around the pillar. The current, $I$, flowing through the dot is measured as a function of gate
voltage, $V_{g}$, in response to a dc voltage, $V$, applied between the
source and drain. (b) Schematic diagram of two single-particle states with
energies $E_{a}$ and $E_{b}$ crossing each other at a magnetic field $B=B_{0}
$. (c) Electrochemical potential, $\mu _{i}(2)=U_{i}(2)-U(1)$, for two
interacting electrons. The thick line depicts the ground state energy
whereas the thin lines show the excited states.}
\label{figure1} 
\end{figure}
The GS has $S=0$ away from $B_{0}$. Near $B_{0}$, the lowest energy is $\mu _{3}(2)$, such
that here $S=1$. The downward cusp in the thick line identifies this
spin-triplet region. The transition in the GS from $S=0$ to 1 and $S$ $=1$
to 0, respectively, occurs when $\mu 1=\mu 3$ for $B$ $<B_{0}$ and when $\mu
_{2}=\mu _{3}$ for $B$ $>$ $B_{0}$. We define two energies, $\Delta _{1}$
and $\Delta _{2}$, to characterize the size of the downward cusp in the GS
at $B$ $=B_{0}$: $\Delta _{1}=\mu _{1}-\mu _{3}=C_{aa}-C_{ab}+|K_{ab}|$, $%
\Delta _{2}=\mu _{2}-\mu _{3}=C_{bb}-C_{ab}+|K_{ab}|$, $\Delta _{1}-\Delta
_{2}=C_{aa}-C_{bb}$.  
Our semiconductor quantum dot (see Fig. 1(a)) has the shape of a
two-dimensional disk [5]. For detecting the GSs, we set the source-drain
voltage, $V$, to a small value. The GSs and excited states (ESs) are both
measured when $V$ is set to a value sufficiently greater than the excitation
energy [6]. For small $V$, a series of current peaks results from changing
the number of electrons in the dot, $N$, one-by-one [1]. The position of a
current peak for the transition from $N-1$ to $N$ measures the GS
electrochemical potential $\mu $($N$). The sample is cooled down to about
100 mK. 
Fig. 2(a) shows the evolution of current peaks with magnetic field for $N=7$
to 16. 
\begin{figure}[b] 
\centerline{\psfig{figure=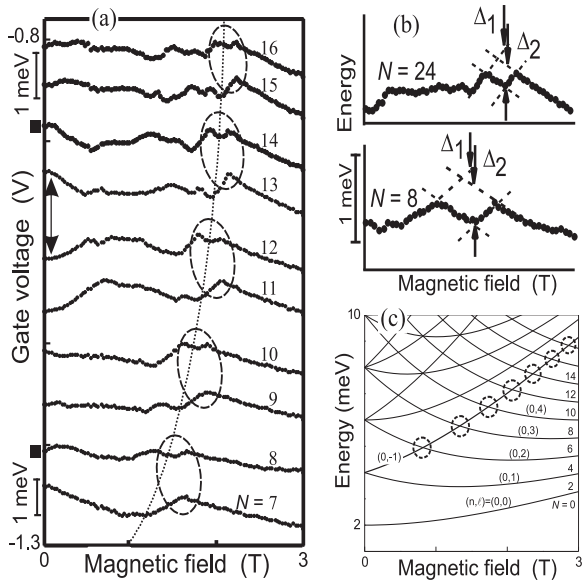,width=11cm,height=11cm,rheight=8.cm}}  
\caption{(a) Evolution of the ground state energies from $N=7$ to 16 as measured from
the current peaks versus magnetic field at $V=120$ mV. The bars along the
gate voltage axis show 1 meV energy scales calibrated at -1.26 and -0.85 V.
The dotted curve indicates the last crossing between single-particle states.
Dashed ovals correlate pairs of ground states for odd and even electron
numbers. Spin transitions in the ground states are indicated by $%
\blacksquare $ at $B=0$ T and occur in the ovals for $B\neq 0$ T. (b)
Magnified plots of the $N=8$ and 24 current peaks vs. magnetic field. The
dashed lines illustrate how the interaction-energy parameters, $\Delta _{1}$
and $\Delta _{2}$ are determined. (c) Fock-Darwin single-particle states
calculated for $\omega _{0}=2$ meV.} 
\label{figure2} 
\end{figure}
Features associated with a parabolic confining potential are all
observed such as a shell structure [2]. The large spacing for $N$ = 12 at $%
B=0$ T can be seen in Fig. 2(a) 
(see double arrow) and marks the complete filling of the first three shells.
The pairing between neighboring peaks indicates anti-parallel spin filling
of a single orbital state by two electrons. Modifications to this pairing
are observed for the peaks labeled by $\blacksquare $ at 0 T, and in each of
the dashed ovals connecting pairs of peaks at non-zero field. These are all
signatures of Hund's first rule; i.e. spin-polarized filling. Note that the
Zeeman effect is negligible in this experiment [7]. We show expansions of
the evolution of the $N=8$ and $N=24$ peaks in Fig. 2(b). The downward cusps
are clearly seen. The dashed lines form a parallelogram, from which we
obtain parameters $\Delta _{1}$ and $\Delta _{2}$. 
To compare the two-electron model with larger electron numbers, we assume
that other states are far away in energy so that they can be neglected.
Then, the downward cusps should occur for higher $even$-electron numbers,
whereas they should be absent for $odd$-electron numbers. This is clearly
observed in the ovals in Fig. 2(a). For instance, the $B$-field dependence
of the 9th peak compares well to the thick line in Fig. 1(b) and the $B$%
-field dependence of the 10th peak compares well to the thick line in Fig.
1(c). Other pairs of even and odd numbered peaks show the same behavior.

\begin{figure}
\centerline{\psfig{figure=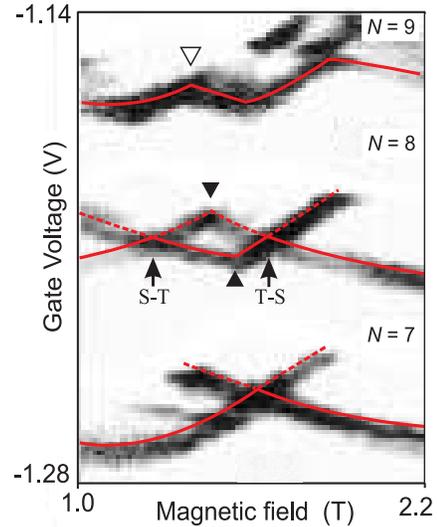,width=8cm,height=10cm,rheight=7.5cm}}
\caption{$dI/dV_{g}$ in the plane of $V_{g}$ and $B$ for $N=7$ to 9 measured for $V=2$
mV. $dI/dV_{g}>0$ for dark blue, $dI/dV_{g}<0$ for white and $dI/dVg\approx 0
$ for yellow. The solid red lines indicate the evolution of the GSs with
magnetic field whereas the dashed red lines show the ESs. The two arrows
indicate singlet-triplet (S-T) and triplet-singlet (T-S) transitions in the
GS for $N=8$.}
\label{figure3}                                                                                      \end{figure}      

This justifies our assumption so that we can simplify the many-electron
system to just one or two electrons. 
More detailed agreement is obtained by measuring the excitation spectrum
[6]. Fig. 3 shows a $dI/dV_{g}$ plot, taken for $V=2$ mV. This larger
voltage opens a sufficiently wide transport window between the Fermi levels
of the source and drain, that both the GS and first few ESs can be detected.
The GS and ESs for $N=7$ to 9 can be assigned from the magnetic field
dependence of the dark blue lines. Solid red lines highlight the GSs whereas
the ESs are indicated by dashed red lines. TThe set of GS and ES lines for $N
$ $=7$ shows a single crossing similar to that in Fig. 1(b). The spectrum
for $N$ $=8$ compares well to Fig. 1(c) and we can clearly distinguish the
parallelogram formed by the GS and first ES. The downward cusp in the GS for 
$N$ $=8$ (labeled $\blacktriangle $) is at a slightly higher $B$-field than
the upward cusp in the first ES (labeled $\blacktriangledown $). This
asymmetry implies that $\Delta _{1}>$ $\Delta _{2}$, i.e. $Caa>Cbb$. The
same type of asymmetry is always observed along the dashed line in Fig.
2(a), implying that $Caa>C_{bb}$ for all $N$. Note that the GS for $N$ $=9$
shows an upward cusp (labeled by $\triangledown $) quite similar in form to
the first ES in the spectrum for $N$ $=8$. This implies that the filling of
the ninth electron is closely linked to the configuration of the $N$ $=8$
first ES [3]. 
As illustrated in Fig. 2(b), we can derive the experimental values for $%
\Delta _{1}$ and $\Delta _{2}$ for different $N$.
These values are plotted in Fig. 4. We find that $\Delta _{1}$ is
larger than $\Delta _{2}$ for all $N$, again implying that $Caa>Cbb$. As $N$
increases from 6 to 12, $\Delta _{1}$ first increases and then slowly
decreases, whilst $\Delta _{2}$ slightly decreases. 

\begin{figure}[b] 
\centerline{\psfig{figure=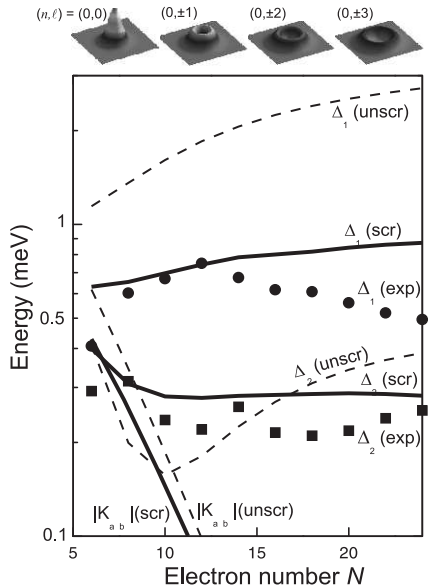,width=8cm,height=11cm,rheight=8cm}}  
\caption{Experimental values for the energy parameters $\Delta _{1}$ ($\bullet $) and $%
\Delta _{2}$ ($\blacksquare $), on a log. scale, versus electron number
derived from data as shown in Fig. 2(a). The uncertainty in the determination of the eperimental values is 10% or less. The dashed and solid curves are
calculated from the FD-wavefunctions at 2 T, for an unscreened
(unscr) and screened (scr) Coulomb interaction The calculated exchange energy 
%TCIMACRO{\TEXTsymbol{\vert}}%
%BeginExpansion
\mbox{$\vert$}%
%EndExpansion
$K_{ab}$%
%TCIMACRO{\TEXTsymbol{\vert} }%
%BeginExpansion
\mbox{$\vert$}%
%EndExpansion
between states with energies $E_{a}=E_{0,-1}$ and $E_{b}=E_{0,N/2-1}$
decreases quickly with $N$. Above the main figure the absolute squares of
the wavefunctions are shown for the relevant quantum numbers $n=0$ and $%
l=0,\pm 1,\pm 2,\pm 3$. As the angular momentum quantum number $l$
increases, the average radius increases.}
\label{figure4} 
\end{figure}

To calculate DC and EX energies [4] we now need to specify the confining
potential of the disk. In earlier work it was shown that the lateral
confinement is well described by a parabolic potential with cylindrical
symmetry [2]. The eigenfunctions with eigenenergies, $E_{n,l}$, in this
potential are known as Fock-Darwin (FD) states [8]: 
\begin{equation}
E_{n,l}=-\frac{l}{2}\hbar \omega _{c}+(n+\frac{1}{2}+\frac{1}{2}|l|)\hbar 
\sqrt{4\omega _{0}^{2}+\omega _{c}^{2}}  \eqnum{1}
\end{equation}
where $n=0,1,2,...$ is the radial quantum number and $l=0,\pm 1,\pm 2,...$
is the quantum number for angular momentum. $\hbar \omega _{0}$ is the
lateral confining energy and $\hbar \omega _{c}$ $=eB/m^{*}$ is the
cyclotron energy. Each FD-state is spin degenerate. At $B=0$ T the
FD-spectrum has sets of states with increasing degeneracy (see Fig. 2(c)).
This degeneracy is lifted on increasing $B$, but as $B$ is increased further
new crossings can occur. The last crossing is always a crossing between just
two FD-states. The up-going state is always $(n,l)=(0,-1)$, whereas the
down-going state, $(0,l>1)$, has an increasing angular momentum for states
with increasing energy. (The relation with Fig. 1 is: $E_{a}$ $=E_{0,-1}$
and $E_{b}=E_{0,l>1}$.) Note that the last crossings also correspond to the
dashed line in Fig. 2(a). 
From the electron distributions of the FD-states we calculate the DC and EX
energies for two electrons occupying two degenerate states. We take $\hbar
\omega _{0}$ $=2$ meV as deduced from earlier experiments [2,6] and obtain $%
\Delta _{1}$ and $\Delta _{2}$. The dashed curves in Fig. 4 show $\Delta _{1}
$ and $\Delta _{2}$ when we neglect screening of the interactions within the
dot by electrons in the leads and in the gate. In this case the Coulomb
potential falls off as $1/r$, where $r$ is the distance between the
electrons [4]. For the solid curves we have approximated the screening
effects by replacing the Coulomb potential by exp$\{-r/d\}/r$. We have taken 
$d$ $=10$ nm which is roughly the thickness of the tunnel barriers. Fig. 4
shows that screening considerably reduces $\Delta _{1}$ to values much
closer to the experimental values. Screening also removes the minimum in $%
\Delta _{2}$, which is also in better agreement with the experiment. Since
the average radius of the wavefunctions increases with angular momentum, two
electrons are closer together when they both occupy (0,-1) compared to when
they both occupy ($0,l=N/2-1$) for even $N$ $>4$ (or $l>1$), so the DC
interaction is stronger in the former. This explains our observation ($%
\Delta _{1}>$ $\Delta _{2}$) $C_{aa}>C_{bb}$ for all $N$. The overlap
between $different$ wavefunctions, (0,-1) and ($0,l=N/2-1$), decreases for
even $N$ $>4$ (or $l>1$). This results in a decrease in both $C_{ab}$ and $%
|K_{ab}|$ with $N$. It then follows that $\Delta _{1}$ increases until it
saturates at a value equal to $C_{aa}$. The gradual decrease of experimental 
$\Delta _{1}$ for $N$ $>12$ is probably related to the decrease in the
lateral confinement with $N$ [2] and thus the decrease in $C_{aa}$.
We finally discuss the interaction effects for the $N$ $=4$, 8 and 14 peaks
near $B$ $=0$ T (the $N=8$ and 14 peaks are labeled n in Fig. 2(a)). These
correspond to the GS electrochemical potentials for adding the second
electron to the second, third and fourth shells, respectively. The inset to
Fig. 5 demonstrates the resemblance to the model of Fig. 1(c) for $N=8$ near 
$B$ $=0$ T. Comparing these data to the FD-spectrum, we assign the states
such that: $E_{a}$ $=E_{0,-2}$ and $E_{b}$ $=E_{0,2}$. Likewise, for $N$ $=4$
we have $E_{a}$ $=$ $E_{0,-1}$ and $E_{b}$ $=$ $E_{0,1}$ [2] and for $N$ $=$
14 we have $E_{a}=E_{0,-3}$ and $E_{b}=E_{0,3}$ [9]. Note that these states
correspond to wavefunctions with a complete overlap. Also, for $B$ $=B_{0}=0$
T the two crossing states have the same orbital symmetry implying $\Delta
_{1}=\Delta _{2}=|K_{ab}|$; i.e. only EX effects contribute to the downward
cusp [10].

\begin{figure}[hbt] 
\centerline{\psfig{figure=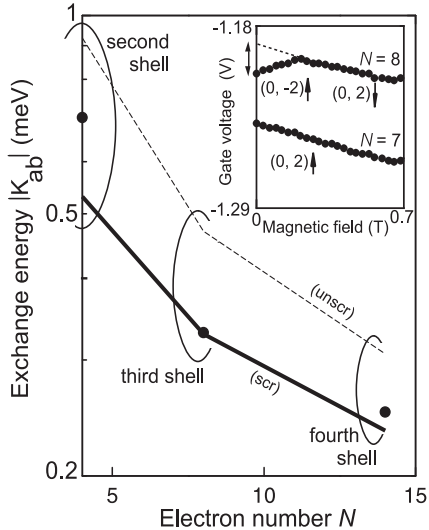,width=8cm,height=10cm,rheight=7.5cm}}  
\caption{Exchange energy $|K_{ab}|$, on a log. scale, associated with spin-triplets
formed when each new shell is filled by just two electrons at $B=0$ T. The
solid circles are the experimental values whose uncertainty in the determination is 10% or less. The inset shows an expansion for
the filling of the first two electrons into the third shell (i.e. $N=7$ and
8). The vertical double-arrow represents $|K_{ab}|$ in units of gate voltage
which is then converted to energy. The calculated curves in the main figure
are for the unscreened (dashed) and screened (solid) cases.}
\label{figure5} 
\end{figure}

We derive 
%TCIMACRO{\TEXTsymbol{\vert}}
%BeginExpansion
\mbox{$\vert$}%
%EndExpansion
$K_{ab}$%
%TCIMACRO{\TEXTsymbol{\vert} }
%BeginExpansion
\mbox{$\vert$}%
%EndExpansion
as illustrated in the inset to Fig. 5. The obtained EX energy quickly
becomes smaller for higher lying shells. For comparison we also show the
calculated screened and unscreened values. The screened case provides the
best quantitative agreement for our realistic choices of the confining
energy and the screening distance. 
Our general model provides a clear identification of effects due to EX and
DC interactions. More advanced calculations support our analyses [11]. An
important simplification is the reduction of a many-electron system to just
two interacting electrons. The type of spin filling in many nearly
degenerate levels near $B$ $=0$ T in larger electron boxes remains an
interesting open issue. We thank G. Bauer, M. Danoesastro, M. Eto, R. van
der Hage, T. Honda, J. Janssen, T. Oosterkamp, H. Tamura, and T. Uesugi for
their help and useful discussions. S. T. and L.P.K. acknowledge financial
support from the Specially Promoted Research, Grant-in-Aid for Scientific
Research, from the Ministry of Education, Science and Culture in Japan, from
the Dutch Organization FOM, from the NEDO program NTDP-98, and from the EU
via a TMR network.

\widetext %need to end this multicol stuff before the end of the document


\begin{references}
\bibitem{4}  $C_{ij}$ and $K_{ij}$ are the usual wavefunction overlap
integrals for an unscreened Coulomb potential:

\begin{eqnarray*}
C_{ij} &=&\int \!\!\!\!\int \left| \psi _{i}(\overrightarrow{r_{1}})\right|
^{2}\frac{e^{2}}{4\pi \varepsilon _{0}\left| \overrightarrow{r_{1}}-%
\overrightarrow{r_{2}}\right| }\left| \psi _{j}(\overrightarrow{r_{2}}%
)\right| ^{2}d\overrightarrow{r_{1}}\overrightarrow{r_{2}} \\
K_{ij} &=&\int \!\!\!\!\int \psi _{i}^{*}(\overrightarrow{r_{1}})\psi _{j}(%
\overrightarrow{r_{1}})\frac{e^{2}}{4\pi \varepsilon _{0}\left| 
\overrightarrow{r_{1}}-\overrightarrow{r_{2}}\right| }\psi _{j}^{*}(%
\overrightarrow{r_{2}})\psi _{i}(\overrightarrow{r_{2}})d\overrightarrow{%
r_{1}}\overrightarrow{r_{2}}
\end{eqnarray*}
$E_{i}$ and $E_{j}$ are eigenvalues for the eigenfunctions $\psi _{i}$ and $%
\psi _{j}$, respectively, and $\overrightarrow{r_{1}}$ and $\overrightarrow{%
r_{2}}$ represent the positions of the two electrons.
\bibitem{9}  For $N$ $=$14 it is difficult to identify the corresponding
state since the $B$-dependence is not clear enough to distinguish between $%
E_{0,-3}$ and $E_{1,-1}$. We assume it is $E_{0,-3}$, and continue to
evaluate $|K_{ab}|$.
\bibitem{10}  Note that the asymmetry ($\Delta _{1}\neq \Delta _{2}$) can
only occur for $B_{0}\neq 0$. Symmetry is required if $B_{0}=0$, implying
that $\Delta _{1}=\Delta _{2}$. Then, the eigenfunctions for the states with
energy $E_{a}$ and $E_{b}$ must have the same spatial form, from which we
conclude that: $C_{aa}=C_{bb}=C_{ab}$ and thus $\Delta _{1}=\Delta
_{2}=|K_{ab}|$.
\end{references}
\end{document}